# Behavior of Dynamical Systems in the Regime of Transient Chaos

**G. B. Astaf'ev, A. A. Koronovskiĭ, and A. E. Hramov**
*State Scientific Center "College," Saratov State University, Saratov, Russia*

**Abstract**—The transient chaos regime in a two-dimensional system with discrete time (Eno map) is considered. It is demonstrated that a time series corresponding to this regime differs from a chaotic series constructed for close values of the control parameters by the presence of "nonregular" regions, the number of which increases with the critical parameter. A possible mechanism of this effect is discussed.

The phenomenon of transient chaos accompanying the crisis of a strange attractor [1] is rather frequently encountered in dynamical systems of various natures. In particular, such phenomena were observed in hydrodynamic experiments [2], radio circuits [3], chemical reactions [4], optical bistable media [5], distributed systems of the type of an electron flow interacting with a backward wave [6], and standard models of nonlinear dynamics such as logistic maps [7], Eno map [8], and Rössler system [9].

The transient chaos is essentially a regime, whereby the distance between a strange attractor and the boundary of its basin of attraction in the phase space decreases until they touch each other at a critical value of the control parameter ($p = p_c$). At this point, the chaotic attractor exhibits a crisis (and ceases to exist at $p > p_c$), converting into an unstable chaotic manifold called chaotic saddle. Accordingly, the behavior of the given system changes, the initial chaotic dynamics being replaced by the transient chaos. In the initial stage of
1 this regime, the system behavior is virtually undistinguishable from chaotic, but then the system rapidly passes to another stable state (attractor) that can be stationary, periodic, or chaotic as well.

A chaotic saddle replacing the chaotic attractor in the phase space is of considerable basic interest. Some properties of the chaotic saddle can be derived from an analysis of the initial stage reflecting the transient chaos regime in a large number of time series [9, 10]. An alternative approach consists in constructing a long artificial time series by matching the initial and end parts of time series, linking them into one [11]. Finally, trajectories existing for a long time in the vicinity of a chaotic saddle can be constructed using the so-called proper interior maximum (PIM) triple [12] and PIM simplex [13] methods, or the method of step perturbations [14]. The resulting long time series offers a good approximation of the chaotic trajectory belonging to the given unstable chaotic manifold.

The aim of this study was to compare the chaotic trajectories obtained for two close values of the control parameter $p$, one corresponding to the chaotic attractor ($p < p_c$) and the other, to the chaotic saddle ($p > p_c$).

The study s performed on an Eno map [15] of the type

$$\begin{aligned} x_{n+1} &= \lambda x_n(1 - x_n) + b y_n, \\ y_{n+1} &= x_n, \end{aligned} \qquad (1)$$

which is a standard model in nonlinear dynamics. For the control parameters $\lambda \approx 4.21934$ and $b = 0.61$, a chaotic attractor based on the cycle of period 3 exhibits a boundary crisis via a series of period doubling bifurcations, loses stability, and transforms into a chaotic saddle. Simultaneously, a stable rest point, representing an attractor, exists in the phase space of this system.

A time series corresponding to the chaotic attractor was constructed for $\lambda = 4.219$ (Fig. 1a). An analogous time series corresponding to the chaotic saddle ($\lambda$ = 4.22 was constructed by the method of step perturbations [14]. (Fig. 1a). This procedure reduces to determining, for an arbitrary point $\mathbf{x}$ in the vicinity of a chaotic saddle, a time $T(\mathbf{x})$ for which a phase trajectory starting from this point reaches the attractor [16]. If the time $T(\mathbf{x})$ proves to be greater than a certain threshold value $T^*$, the point $\mathbf{x}$ is considered as belonging to an unstable chaotic manifold. In order to construct a long time series corresponding to the regime of transient chaos, the initial dynamical system is iterated as follows:

$$\mathbf{x}_{n+1} = \begin{cases} \mathbf{F}(\mathbf{x}_n) & \text{for} \quad T(\mathbf{x}_n) > T^*(\text{step}), \\ \mathbf{F}(\mathbf{x}_n + \mathbf{r}_n) & \text{for} \quad T(\mathbf{x}_n) \le T^*(\mathbf{r}_n \text{ is perturbation}), \end{cases} \qquad (2)$$

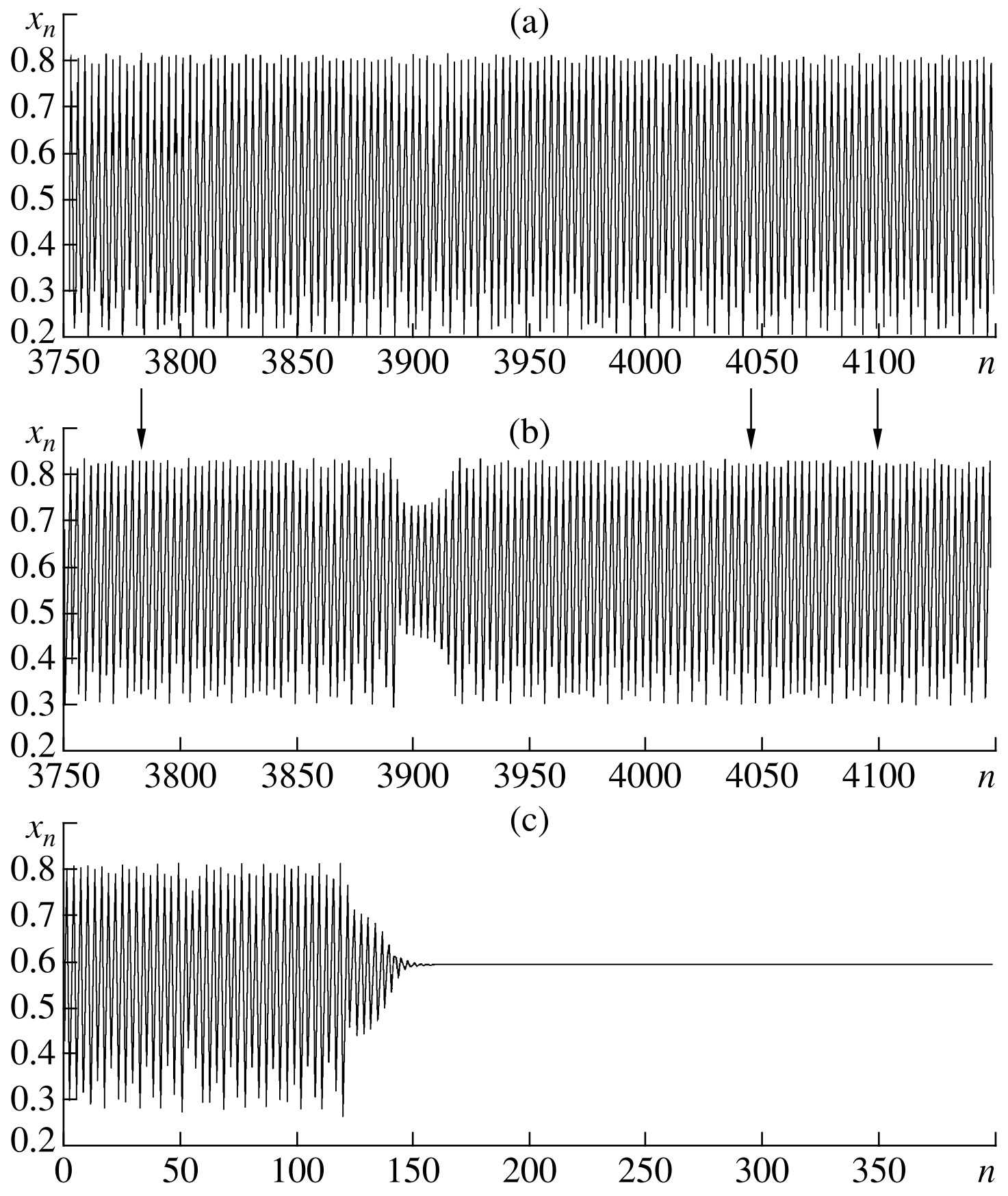

**Fig. 1.** Fragments of time series for the Eno maps (1) with $b = -0.61$: (a) exhibiting a chaotic attractor ($\lambda = 4.219$); (b) featuring a chaotic saddle ($\lambda = 4.22$); (c) escaping from the transient chaos regime ($\lambda = 4.22$). The time series (b) were constructed by the method of step perturbations (arrows indicate the moments of discrete time $n$ corresponding to the introduction of small perturbations $\mathbf{r}$).

where $\mathbf{F}(\mathbf{x})$ is the operator of evolution of the initial dynamical system, $\mathbf{r}_n$ is a perturbation vector such that $T(\mathbf{x}_n + \mathbf{r}_n) > T^*$, and $|\mathbf{r}_n| < \delta = 10^{-7}$. Here, vector $\mathbf{r}_n$ is a random quantity but, in order to accelerate construction of the time series, the probability density distribution $p(\mathbf{r})$ has to obey certain requirements (see [14]).

Thus, until the time required for an imaging point to attain the attractor exceeds a certain preset (larger large) threshold $T^*$, the point is considered as occurring near the chaotic saddle. The initial dynamical system, representing the Eno map (1), was iterated according to this procedure. As soon as the time of attaining the attractor becomes equal to the threshold value, the point on the trajectory is variously perturbed and the possible perturbation $\mathbf{r}_n$ is selected for which the time $T(\mathbf{x}_n + \mathbf{r}_n)$ required for the perturbed imaging point to attain the attractor would exceed the threshold $T^*$. Since the perturbation vector $\mathbf{r}$ is small ($|\mathbf{r}_n| < \delta = 10^{-7}$), the sequence of $\mathbf{x}_n$ offers a good approximation to the chaotic saddle.

According to the results of our investigations, a time series corresponding to the regime of transient chaos differs from a chaotic series only by the presence of "outbursts" representing nonregular regions in which the system dynamics differs from typical. The rest of the time, the chaotic time series is virtually identical to that of the transient chaos (Fig. 1). It should be noted that the appearance of nonregular regions in the time series of transient chaos is by no means related to the small perturbations $\mathbf{r}$ introduced into the system dynamics. In Fig. 1b, the moments of discrete time $n$ corresponding to the introduction of perturbations are indicated by the arrows. As can be seen, the system shows typical behavior for a rather long time after the moment of perturbation and then features an untypical region in the time series. Moreover, by no means each perturbation is followed by such an untypical response.

As the critical parameter $\lambda - \lambda_c$ increases, the number of nonregular regions grows whereas regular regions become shorter. This system behavior in the transient chaos regime is much like intermittency [17, 18], whereby a time series exhibits alternating laminar and turbulent parts, or the crisis induced intermittency [19], in which case several regions of a chaotic attractor merge upon a crisis and the imaging point, having spent some time in one region of the attractor, jumps to another region of this attractor. The appearance of such nonregular regions in the time series corresponding to the transient chaos regime is caused much by the same factors as the escape of the imaging point from a chaotic saddle region [20].

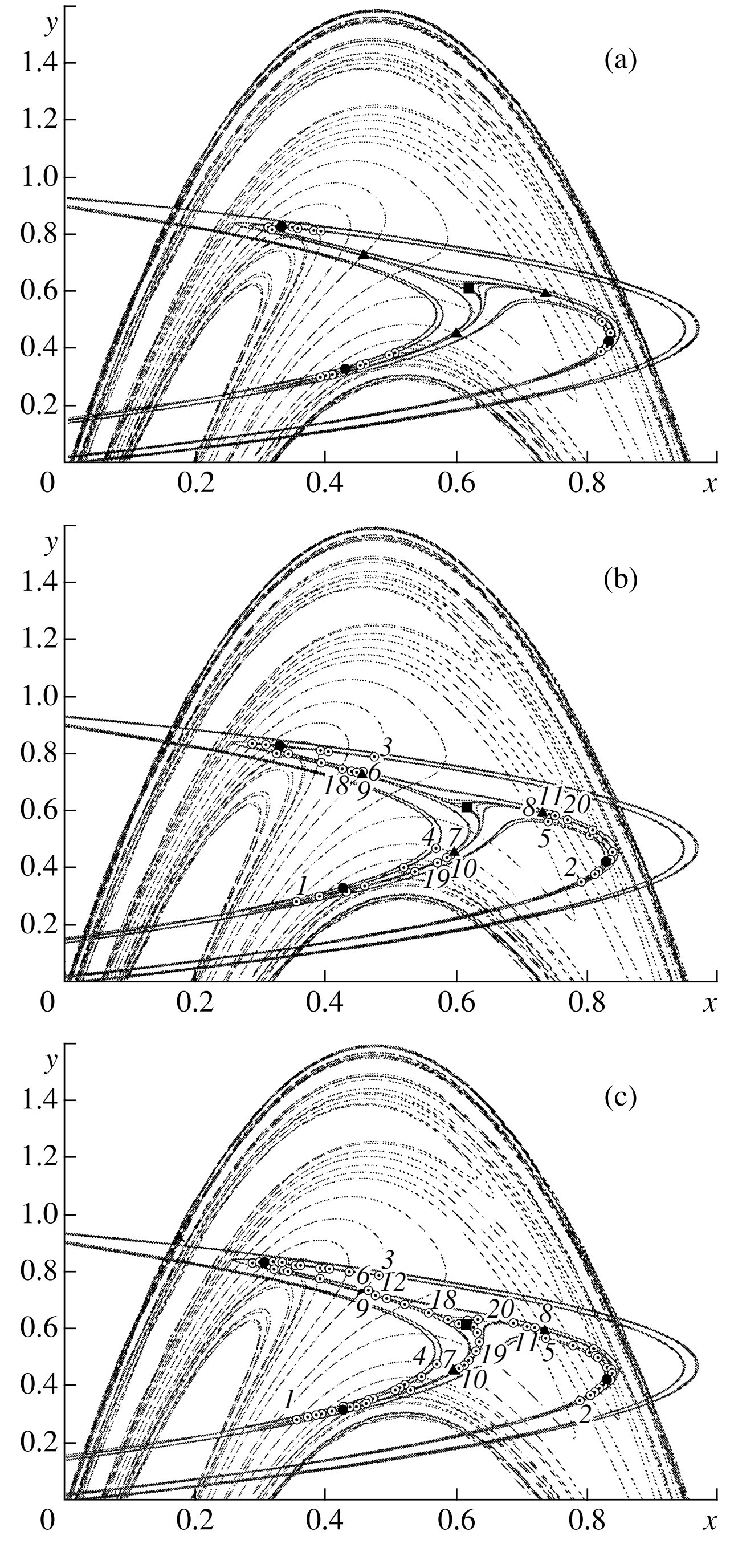

**Fig. 2.** Schematic diagrams showing the positions of a stable point (■), unstable cycles of period 3 (▲ and ○), and the unstable (solid curves) and stable (dashed curves) manifolds of these cycles for the Eno map (1) with $\lambda = 4.22$ and $b = -0.61$: (a) points (○) in a typical region of the time series corresponding to the transient chaos regime; (b) points in the untypical region (figures indicate the numbers of iteration steps, beginning with the first, in which the imaging point falls in the region of intersection of the stable and unstable manifolds of the period-3 cycle, which accounts for the appearance of the untypical region); (c) points in the region where the system escapes from the transient chaos regime (figures indicate the numbers of iteration steps, beginning with the first, in which the imaging point falls in the region of intersection of the stable and unstable manifolds of the period-3 cycle, which accounts for the escape).

Figure 2 shows the structure of manifolds for two unstable cycles of period 3, which coexist in the phase space with an immobile stable point in a system with the control parameters $\lambda = 4.22$ and $b = -0.61$. The manifolds were mapped using the method described in [21–23]. As can be seen, stable and unstable manifolds exhibit a rather complicated structure, intersecting each other an infinite number of times. Figure 2a shows a trajectory of the imaging point for the system exhibiting typical behavior. Here, the points of a chaotic saddle are arranged along unstable manifolds of the unstable cycles of period 3 and are confined within a certain region.

An untypical region in the time series begins at the moment when the imaging point, moving in the vicinity of the unstable manifold of the unstable cycle of period 3, falls in the vicinity of the region of intersection of the unstable and stable manifolds of this cycle (point *1* in Fig. 2b). Then, the imaging point begins to visit regions in the $(x, y)$ phase plane in which the stable and unstable manifolds intersect, thus approaching the unstable cycle of period 3. After entering the vicinity of this cycle, the imaging point occurs for some time in this region and then returns (along the unstable manifold of the unstable cycle) to the chaotic saddle region (Fig. 2b). The return of the imaging point to this region corresponds to completion of an untypical region in the time series.

An analogous mechanism accounts for escape of the system from the transient chaos regime (Fig. 2c). The only difference consists in that the imaging point enters the vicinity of the unstable cycle of period 3 three from the other side of a relatively stable manifold. Then, moving along an unstable manifold of the unstable cycle for a relatively short period of discrete time, the imaging point reaches an attractor (in this example, an immobile point). In this sense, the stable manifold of the unstable cycle is essentially a boundary separating regions of a fast transient process and transient chaos on the phase plane. This is similar to the case of multistability, where a stable manifold of the unstable cycle is a boundary of the basins of attraction of the attractors [24, 25]. To summarize, we have demonstrated that time series of a dynamical system with discrete time, occurring in the transient chaos regime, contain regions of untypical behavior reflecting approach of the imaging point to an unstable cycle along a stable manifold, followed by return of the imaging to the chaotic saddle region.

**Acknowledgments.** This study was supported by the Russian Foundation for Basic Research (project no. 02-02-16351) and by the Scientific-Education Center “Nonlinear Dynamics and Biophysics” at the Saratov State University (Grant REC-006 from the U.S.